\documentstyle[prd,aps,preprint,psfig]{revtex}

\newcommand{\CL}{{\cal L}}

\newcommand{\CO}{{\cal O}}

\newcommand{\bear}{\begin{array}}  \newcommand{\eear}{\end{array}}
\newcommand{\bea}{\begin{eqnarray}}  \newcommand{\eea}{\end{eqnarray}}
\newcommand{\beq}{\begin{equation}}  \newcommand{\eeq}{\end{equation}}
\newcommand{\bef}{\begin{figure}}  \newcommand{\eef}{\end{figure}}
\newcommand{\bec}{\begin{center}}  \newcommand{\eec}{\end{center}}
\newcommand{\non}{\nonumber}  
\newcommand{\lmk}{\left(}  \newcommand{\rmk}{\right)}
\newcommand{\lkk}{\left[}  \newcommand{\rkk}{\right]}
\newcommand{\lhk}{\left \{ }  \newcommand{\rhk}{\right \} }
  
\newcommand{\lla}{\left \langle }  \newcommand{\rra}{\right \rangle }
\newcommand{\del}{\partial}  

\newcommand{\bib}{\bibitem}

\def\IB#1#2#3{{\bf #1}, #2 (19#3)}
\def\IBID#1#2#3{{\it ibid}. {\bf #1}, #2 (19#3)}

\def\NATT#1#2#3{Nature (London) {\bf #1}, #2 (20#3)}
\def\NPB#1#2#3{Nucl. Phys. {\bf B#1}, #2 (19#3)}

\def\PLB#1#2#3{Phys. Lett. B {\bf #1}, #2 (19#3)}
\def\PLBold#1#2#3{Phys. Lett. {\bf#1B}, #2 (19#3)}

\def\PRD#1#2#3{Phys. Rev. D {\bf #1}, #2 (19#3)}
\def\PRDD#1#2#3{Phys. Rev. D {\bf #1}, #2 (20#3)}

\def\PRL#1#2#3{Phys. Rev. Lett. {\bf#1}, #2 (19#3)}
\def\PRLL#1#2#3{Phys. Rev. Lett. {\bf#1}, #2 (20#3)}
\def\PRT#1#2#3{Phys. Rep. {\bf#1}, #2 (19#3)}

\def\MPL#1#2#3{Mod. Phys. Lett. A {\bf #1}, #2 (19#3)}

\def\APJL#1#2#3{Astrophys. J. Lett. {\bf #1}, L#2 (19#3)}
\def\APJLL#1#2#3{Astrophys. J. Lett. {\bf #1}, L#2 (20#3)}

\def\CQG#1#2#3{Class. Quantum Grav. {\bf #1}, #2 (19#3)}
\def\PTP#1#2#3{Prog. Theor. Phys. {\bf #1}, #2 (19#3)}

\def\MNRAS#1#2#3{Mon. Not. R. Astron. Soc. {\bf #1}, #2 (19#3)}

\begin{document}

\tighten
\draft
\title{New inflation in supergravity with a chaotic initial condition}

\author{Masahide Yamaguchi}
\address{Research Center for the Early Universe, University of Tokyo,
  Tokyo 113-0033, Japan}
\author{Jun'ichi Yokoyama}
\address{Department of Earth and Space Science, Graduate School of
Science, Osaka University, Toyonaka 560-0043, Japan}

\date{\today}

\maketitle

\begin{abstract}
    We propose a self-consistent scenario of new inflation in
    supergravity. Chaotic inflation first takes place around the
    Planck scale, which solves the longevity problem, namely, why the
    universe can live much beyond the Planck time, and also gives an
    adequate initial condition for new inflation. Then, new inflation
    lasts long enough to generate primordial fluctuations for the
    large scale structure, which generally has a tilted spectrum with
    the spectral index $n_{s} < 1$. The successive decay of the
    inflaton leads to a reheating temperature low enough to avoid the
    overproduction of gravitinos in a wide range of the gravitino
    mass.
\end{abstract}



\section{Introduction}

\label{sec:introduction}

Inflation is the most natural extension of standard big-bang cosmology
because it not only solves the flatness and the horizon problems but
also provides a generation mechanism of density fluctuations
\cite{oriinf,newinf,inflation}. Since inflation typically takes place
at an energy scale much higher than the electroweak scale, most
favorably around the Planck scale in fact, we cannot but seriously
confront the hierarchy problem in particle physics.  To this end, it
is inevitable to consider inflation models in supersymmetric theory,
in particular, its local version, supergravity~(SUGRA) \cite{SUSY,LR}.

Among the various mechanisms of inflation proposed so far, chaotic
inflation \cite{chaoinf} is the most attractive in that it starts
around the Planck scale so that it does not suffer from the longevity
problem, that is, why the universe lives beyond the Planck time.
However, it has been recognized that it is difficult to realize
chaotic inflation in SUGRA because scalar potentials in minimal SUGRA
have an exponential factor of the form $\exp(|\phi|^2/M_G^2)$ which
forbids any scalar field $\phi$ to have a value much larger than the
reduced Planck scale $M_G\simeq 2.4\times 10^{18}$ GeV and hampers
chaotic inflation.  Although several supergravity chaotic inflation
models have been proposed so far using functional degrees of freedom
in SUGRA as a nonrenormalizable theory \cite{GL,MSY2}, these models
employ rather specific K\"ahler potentials and require fine tuning
since there are no symmetry reasons for having the proposed forms.
Recently, however, Kawasaki, Yanagida, and one of the
authors~(M.Y.)\cite{KYY} proposed a natural model of chaotic inflation
in SUGRA using the Nambu-Goldstone-like shift symmetry. But, in this
model, the reheat temperature low enough to avoid overproduction of
gravitinos is realized taking a rather small coupling constant $(\sim
10^{-5})$ though it is natural in the 't Hooft's sense \cite{tHooft}.

As a model of inflation that predicts low reheating temperature
straightforwardly \cite{KMY,IY,DR}, new inflation is more attractive
than chaotic inflation because it occurs at a lower energy scale.
Furthermore, new inflation can easily generate density fluctuations
\cite{fluc} with a tilted spectrum, which may naturally explain the
recent observation of anisotropies of the cosmic microwave background
radiation (CMB) by the BOOMERANG experiment \cite{BOOMERANG} and the
MAXIMA experiment \cite{MAXIMA}. However, it suffers from a severe
problem about the initial value of the inflaton \cite{inflation} in
addition to the longevity problem mentioned above. In order to realize
successful inflation, the initial value of the inflaton must be
fine-tuned near the local maximum of the potential over the horizon
scale.  For this problem Asaka, Kawasaki, and one of the
authors~(M.Y.)\cite{AKY} proposed a solution by considering the
gravitationally suppressed interactions with particles in the thermal
bath. But the longevity problem remains.  Izawa, Kawasaki, and
Yanagida \cite{IKY}, on the other hand, considered another inflation
(called pre-inflation) which takes place before new inflation and
drives the scalar field responsible for new inflation dynamically
toward the local maximum of its potential. If the preinflation is
chaotic inflation, the longevity problem is solved, too.

Thus we are naturally motivated to a model of successive inflation
\cite{KLS}, namely, chaotic inflation followed by new inflation.  In
fact, such a double inflation model has already been proposed in a
different context \cite{chaonew}, but in this paper, we propose a
simple and self-consistent model of successive inflation in the
framework of SUGRA in which the two inflatons belong to the same
supermultiplet.  That is, the inflaton for chaotic inflation is the
imaginary part of a complex scalar field while its real part drives
new inflation.  In fact, our model is a triple inflation model where
chaotic inflation first takes place followed by a mini inflation
driven by a false vacuum energy and then new inflation occurs.

\section{Model}

\label{sec:model}

We introduce the inflaton chiral superfield $\Phi(x,\theta)$ and
assume that the K\"ahler potential, $K(\Phi,\Phi^{\ast},\dots)$, is a
function of $\Phi+\Phi^{\ast}$, so that the imaginary part of the
scalar component of the superfield $\Phi$, $\chi$, can take a value
larger than the Planck scale without costing exponentially large
potential energy density and that it may be identified with the
inflaton for chaotic inflation as below.

Such a functional dependence of $K$ may be accounted for by imposing
the shift symmetry
\beq
  \Phi \rightarrow \Phi + i~C M_{G},
  \label{eq:shift}
\eeq
where $C$ is a dimensionless real constant and $M_G$ is the reduced
Planck scale which we take to be unity hereafter \cite{KYY}. If this
Nambu-Goldstone-like symmetry is exact, however, $\chi$ cannot have
any potential and chaotic inflation would be impossible no matter how
large an initial value it could take.  Hence this global symmetry must
be explicitly broken.  In fact, it has been argued in the literature
\cite{break} that global symmetries may inevitably be broken due to
quantum gravity effects, which may help to generate a potential to the
inflaton.

On the other hand, since the theory of quantum gravity is far from
established yet, it is by no means possible to calculate the
corrections due to quantum gravity explicitly.  Hence, in this paper,
we simply introduce the following superpotential which breaks the
shift symmetry with a dimensionless coupling parameter $g'$,
\beq
  W = v^{2} X - g' X \Phi^{2} = v^{2}X(1-g\Phi^{2}),
\eeq
where we have introduced another chiral superfield $X(x,\theta)$.
Here $v$ is a scale generated dynamically, and $g\equiv g'v^{-2}$.

One may be tempted to add higher-order terms that may be induced by
quantum gravity.  Chaotic inflation is still possible even after these
correction terms are taken into account, provided that the scalar
potential is not steeper than an exponential function and that the
inflaton can take a value several times larger than the Planck scale
without exceeding the Planckian energy density \cite{inflation}. We
assume quantum gravity corrections respect these properties. We also
omit these uncalculable higher-order terms below, for their effects
would become unimportant well before chaotic inflation ends when the
energy density is much smaller than the Planck scale.

The above superpotential $W$ is invariant under $U(1)_{R} \times
Z_{2}$ symmetry. Under $U(1)_{R}$ symmetry,
\bea
    X(\theta) &\rightarrow& e^{-2i\alpha} X(\theta e^{i\alpha}), \\ 
    \Phi(\theta) &\rightarrow& \Phi(\theta e^{i\alpha}).
\eea  
Also, $X$ is even and $\Phi$ is odd under $Z_{2}$ symmetry. Although
this superpotential $W$ is not invariant under the
Nambu-Goldstone-like shift symmetry, it is natural in the sense of 't
Hooft \cite{tHooft}.  For the symmetry is recovered if the small
parameter $g'$ is set to be zero. Since the correction to the K\"ahler
potential from this breaking term is negligible if $g \ll 1$, we
consider the K\"ahler potential which is invariant under the
Nambu-Goldstone-like shift symmetry and $U(1)_{R} \times Z_{2}$
symmetry,
\beq
  K(\Phi, \Phi^{\ast}, X, X^{\ast}) = 
     K[(\Phi + \Phi^{\ast})^{2}, XX^{\ast}].  
\eeq
Below we take a minimal K\"ahler potential, for which scalar kinetic
terms are canonical,
\beq
  K = \frac12 (\Phi + \Phi^{\ast})^{2} + XX^{\ast},
  \label{eq:minikahler}
\eeq
and extend it later to incorporate higher-order effects.

\section{Dynamics}

\label{sec:dynamics}

The scalar Lagrangian density $L(\Phi,X)$ is given by
\beq
  L(\Phi,X) = \partial_{\mu}\Phi\partial^{\mu}\Phi^{\ast} 
  + \partial_{\mu}X\partial^{\mu}X^{\ast}
         -V(\Phi,X),
\eeq
where the scalar components of the superfields are denoted by the same
symbols as the corresponding superfields.  The scalar potential $V$
of the chiral superfields $X(x,\theta)$ and $\Phi(x,\theta)$ is given
by
\beq
  V = e^{K} \left\{ \left(
      \frac{\partial^2K}{\partial z^{i}\partial z_{j}^{*}}
    \right)^{-1}D_{z^{i}}W D_{z_{j}^{*}}W^{*}
    - 3 |W|^{2}\right\},~~~~~~~(z^i = \Phi, X )
  \label{eq:potential}
\eeq
with 
\beq
  D_{z^i}W = \frac{\partial W}{\partial z^{i}} 
    + \frac{\partial K}{\partial z^{i}}W.
  \label{eq:DW}
\eeq
It is explicitly given by
\beq
  V = v^{4} e^{K} \lkk~\left|1-g\Phi^{2}\right|^{2}(1-|X|^{2}+|X|^{4}) 
     + |X|^{2} \left|-2g\Phi + (\Phi+\Phi^{\ast})(1-g\Phi^{2})~\right|^{2}~\rkk.
\eeq
Now, we decompose the scalar field $\Phi$ into real and imaginary
components,
\beq
  \Phi = \frac{1}{\sqrt{2}} (\varphi + i \chi).
\eeq
Then, the Lagrangian density $L(\varphi,\chi,X)$ is given by
\beq
  L(\varphi,\chi,X) = 
              \frac{1}{2}\partial_{\mu}\varphi\partial^{\mu}\varphi 
              + \frac{1}{2}\partial_{\mu}\chi\partial^{\mu}\chi 
              + \partial_{\mu}X\partial^{\mu}X^{*}
              -V(\varphi,\chi,X),
\eeq
with the potential $V(\varphi,\chi,X)$ given by
\bea
  V(\varphi,\chi,X)
    &=& v^{4} \exp(|X|^{2}+\varphi^{2})
         \lhk~\lkk 
             \lmk 1-\frac{g}{2}\varphi^{2} \rmk^{2} 
             +g\chi^{2} \lmk
                         1+\frac{g}{2}\varphi^{2}+\frac{g}{4}\chi^{2}
                        \rmk
              \rkk
             (1-|X|^{2}+|X|^{4}) 
         \right. \non \\ 
    && \hspace{-2.0cm} \left.
             +|X|^{2} 
              \lkk
             2g^{2}\chi^{2}+2(g-1)^{2}\varphi^{2}
             +2g(g+1)\varphi^{2}\chi^{2}+2g(g-1)\varphi^{4}
             +\frac{g^{2}}{2}\varphi^{2}(\varphi^{2}+\chi^{2})^{2}
              \rkk~
         \rhk.
\eea

While $\varphi, |X| \lesssim \CO(1)$ due to the factor
$e^{|X|^{2}+\varphi^{2}}$, $\chi$ can take a value much larger than
unity without costing exponentially large potential energy. Then, the
amplitudes of both $\varphi$ and $X$ soon become smaller than unity
due to this steep potential and the exponential factor can be Taylor
expanded around the origin.  This is the situation we deal with
hereafter.  Then the scalar potential is dominated by
\beq
  V \simeq \frac14 \lambda \chi^{4}
\eeq
with $\lambda = g^{2} v^{4}$.

Thus chaotic inflation can set out around the Planck epoch.  During
chaotic inflation, the mass squared of $\varphi$, $m_{\varphi}^{2}$,
reads
\beq
  m_{\varphi}^{2} \simeq \frac12 g^{2} v^{4}\chi^{4}
                  \simeq 6H^{2},~~~~~~H^{2} \simeq \frac{\lambda}{12}\chi^{4},
\eeq
where $H$ is the Hubble parameter in this era. Then, $\varphi$ oscillates
rapidly around the origin so that its amplitude damps in proportion to
$a^{-3/2}$ with $a$ being the scale factor.  At the end of chaotic
inflation, the amplitude of $\varphi$, $\varphi_{\rm end}$, reads
\beq
  \varphi_{\rm end} \simeq \varphi_{\rm init} 
                            e^{-(3/2) N_{\rm ch}},
  \label{eq:varphiend}
\eeq
classically, where $\varphi_{\rm init}$ is the value of $\varphi$ in
the beginning of chaotic inflation with its natural magnitude being of
order of unity, and $N_{\rm ch}$ is the number of $e$ folds during
chaotic inflation which is usually very large.  Thus $\varphi$
practically vanishes classically and its magnitude is dominated by
quantum fluctuations.  We therefore find $\varphi_{\rm end} \sim
H_{\rm end} \sim \lambda^{1/2} = gv^2$.

On the other hand, the mass squared of $X$, $m_{X}^{2}$, is dominated
by
\beq
  m_{X}^{2} \simeq 2 g^{2} v^{4} \chi^{2} \simeq \frac{24}{\chi^2}H^2, \label{mhratio}
\eeq
which is much smaller than the Hubble parameter in the early stage of
chaotic inflation, when $X$ moves towards the origin only slowly.
Below we set $X$ to be real and positive making use of the freedom of
the phase choice.  In this regime classical equations of motion for
$X$ and $\chi$ are given, respectively, by
\bea
  3H &\dot{X}& \simeq - m_{X}^{2} X,  \label{Xeq} \\
  3H &\dot{\chi}& \simeq - \lambda \chi^{3}, \label{chieq}
\eea
from which we find
\beq
  X \propto \chi^{2}.   \label{propto}
\eeq
This relation holds actually if and only if quantum fluctuations are
unimportant for both $\chi$ and $X$.  As for $\chi$, the amplitude of
quantum fluctuations acquired in one expansion time is larger than the
magnitude of classical evolution in the same period if $\chi \gtrsim
\lambda^{-1/6}$ \cite{eternal}, when the universe is in a
self-reproduction stage of eternal inflation \cite{eternal,sr}.  Hence
let us consider the regime $\chi \ll \lambda^{-1/6}$ and Eq. (\ref{chieq})
holds. Then we can estimate the root-mean-square (rms) fluctuation in
$X$ using the Fokker-Planck equation for the statistical distribution
function of $X$, $P[X,t]$,
\beq
  \frac{\partial~}{\partial t}P[X,t]
=\frac{1}{3H(t)}\frac{\partial~}{\partial X}\lmk m_X^2XP[X,t]\rmk
+ \frac{H^3(t)}{8\pi^2}\frac{\partial^2~}{\partial X^2}P[X,t], \label{FPeq}
\eeq
which is obtained on the basis of Eq. (\ref{Xeq}) using the stochastic
inflation method of Starobinsky \cite{stochastic}. Time evolution
of the rms fluctuation of $X$ is given by
\beq
  \frac{d}{dt} \lla \lmk\Delta X\rmk^2 \rra = 
     - \frac{2 m_X^2}{3 H} \lla \lmk\Delta X\rmk^2 \rra
      + \frac{H^{3}}{4\pi^{2}}.
  \label{evXfluc}
\eeq
Taking $\chi$ as a time variable in Eq. (\ref{evXfluc}) by virtue of
Eq. (\ref{chieq}), we find that the rms fluctuation of $X$ in an initially
homogeneous domain at $\chi=\chi_i$ is given by
\beq
   \lla \lmk\Delta X\rmk^2 \rra = \frac{\lambda}{384\pi^{2}}
   \lmk \chi_i^2\chi^4 -\chi^6 \rmk   \label{Xfluc}
\eeq
at the epoch $\chi$. Taking $\chi_i \sim \lambda^{-1/6}$, $\lla
\lmk\Delta X\rmk^2 \rra$ asymptotically approaches
\beq
   \lla \lmk\Delta X\rmk^2 \rra \simeq
   \frac{\lambda^{2/3}}{384\pi^{2}}\chi^4,  \label{Xyuragi}
\eeq
which is much less than unity because $\lambda$ must be a tiny number
as will be shown later.  From Eqs. (\ref{propto}) and (\ref{Xyuragi})
the amplitude of $X$ becomes much smaller than unity by the time $\chi
\simeq \sqrt{24}$.  Thereafter Eq. (\ref{Xeq}) no longer holds and $X$
oscillates around the origin rapidly and its amplitude decreases even
more.  Thus our approximation that both $\varphi$ and $X$ are much
smaller than unity is consistent throughout the chaotic inflation
regime.

As $\chi$ becomes of order of unity, chaotic inflation ends and the
field oscillates coherently with the mass squared $m_{\chi}^{2} \simeq
2gv^{4}$.  Since $g$ must take a value slightly larger than unity as
shown later, the energy density of this oscillation becomes less than
the vacuum energy density $\sim v^{4}$ soon.  At this stage $\varphi$
is still localized at the origin since its mass squared is given by
\beq m_{\varphi}^{2} \simeq -(g-1) + g \lmk 1 + \frac{g}2 \rmk
\chi^{2}, \eeq which becomes negative only after the square amplitude
of $\chi$ becomes smaller than $\chi_c^2 \equiv 2(g-1)/3 \ll 1$.
Thus the second inflation takes place, which is driven by a false
vacuum energy, before $\varphi$ becomes unstable.

Once the inflation occurs, the amplitude of the oscillation rapidly
goes to zero because the amplitude damps in proportional to
$a^{-3/2}$. Then, the potential energy with $\chi \simeq 0$ reads
\bea
  V(\varphi) &\simeq& v^{4} \exp \lmk \varphi^{2}+|X|^{2} \rmk \non \\
      && \times
      \lhk~\lmk 1 - \frac{g}{2}\varphi^{2} \rmk^{2}(1-|X|^{2}+|X|^{4})
      +|X|^{2} \lkk
             2(g-1)^{2}\varphi^{2}+2g(g-1)\varphi^{4}
             +\frac{g^{2}}{2}\varphi^{6}
             \rkk~ 
      \rhk \non \\
      &\simeq& v^{4} \lkk 1 - \frac{ c }{2}\varphi^{2} 
             + 2(g-1)^{2}\varphi^{2} |X|^{2} \rkk 
      \qquad \qquad ({\rm for}~\varphi, |X| \ll 1) \non \\
      &\simeq& v^{4} - \frac{ c }{2}v^{4}\varphi^{2},
  \label{eq:newpot}
\eea
with $c \equiv 2(g-1)$. Thus, if $g \gtrsim 1$, $\varphi$ rolls down
slowly toward the vacuum expectation value $\eta = \sqrt{2/g}$ and new
inflation takes place. Here and hereafter we set $\varphi$ to be
positive by use of $Z_{2}$ symmetry ($\Phi \leftrightarrow -\Phi$).

As shown in Eq. (\ref{eq:varphiend}), the initial value of $\varphi$ in
the beginning of new inflation is set by the amplitude of quantum
fluctuation and is of order of $H$ then.  The dynamics of $\varphi$ is
also governed by quantum fluctuations until the classical motion
during the Hubble time, $\Delta\varphi \sim |\dot{\varphi}|H^{-1}$,
becomes larger than the quantum fluctuations, $H/(2\pi)$, acquired
during the same period. In our case, this condition is equivalent to
\beq
  \varphi > \varphi_{c} \equiv \frac{v^{2}}{2\sqrt{3}\pi c}
  =\frac{H}{2\pi c}.
\eeq
Since $ c < 1$ as shown later, quantum fluctuations dominate the
dynamics initially. Therefore the universe enters the
self-regenerating stage \cite{eternal,sr} so that the imprint of
chaotic inflation is washed away and current horizon scale must be
contained in one of a new inflation domain in which $\varphi$ got
larger than $\varphi_c$ and the classical description of the dynamics
of $\varphi$ with the potential Eq. (\ref{eq:newpot}) became possible.
The slow-roll condition for the inflaton $\varphi$ is satisfied for
$0< c < 1$ and $0 \lesssim \varphi \lesssim 1$. The Hubble parameter
during new inflation is given by $H \simeq v^2/\sqrt{3}$. Then, the
number of $e$ folds acquired for $\varphi > \varphi_N$ is given by
\beq
  N \simeq \int_{\varphi_{f}}^{\varphi_{N}} \frac{V}{V'} 
    \simeq -\frac{1}{ c } 
              \ln \lmk \frac{\varphi_{N}}{\varphi_{f}} \rmk 
    \simeq -\frac{1}{ c } \ln \varphi_{N},          
\eeq
where the prime represents the derivative with respect to $\varphi$
and $\varphi_{f} \sim 1$ is the value of $\varphi$ at the end of new
inflation. In the new inflation regime, both $\varphi$ and $X$ acquire
large quantum fluctuations.  Following the same procedure as
Ref. \cite{SY}, however, we can show that only $\varphi$ contributes to
adiabatic fluctuations in the present situation.  Hence the amplitude
of curvature perturbation $\Phi_H$ on the comoving horizon scale at
$\varphi=\varphi_N$ is given by the standard one-field formula and
reads
\beq
  \Phi_H(N) \simeq \frac{f}{2\sqrt{3}\pi} 
  \frac{v^{2}}{ c \varphi_{N}}, 
\eeq
where $f=3/5~(2/3)$ in the matter (radiation) domination.  Since the
Cosmic Background Explorer (COBE) normalization requires $\Phi_H(N)
\simeq 3\times 10^{-5}$ at $N\simeq 60$ \cite{COBE}, the scale $v$ is
given by
\bea
    v  \simeq  2.3 \times 10^{-2} \sqrt{ c } 
    e^{-cN/2}|_{N=60}
    \simeq 1.8 \times 10^{-3} - 3.6 \times 10^{-4}
\eea
for $0.02 \le c \le 0.1$. The spectral index $n_s$ of the density
fluctuations is given by
\beq
    n_s \simeq 1 - 2  c .
\eeq
The data of anisotropies of the cosmic microwave background radiation
(CMB) by the COBE satellite \cite{COBE} also implies $n_s = 1.0 \pm
0.2$ so that $0 < c < 0.1$, which leads to $1.00 < g < 1.05$.

After new inflation, $\varphi$ oscillates around the minimum $\varphi
= \eta$ and the universe is dominated by a coherent scalar-field
oscillation of $\sigma \equiv \varphi -\eta$.  Since the exponential
factor $e^{\varphi^2} $ in $e^K$ can be expanded as
\[
  e^{\varphi^2}= e^{\eta^2}(1+2\eta\sigma+\cdots ),
\]
$\sigma$ has linear interactions of gravitational strength with all 
scalar and spinor fields in the theory including those in the minimal
supersymmetric standard model (MSSM).  Let us consider, for example,
the term $W = y_{i}D_{i}HS_{i}$ in the superpotential in
MSSM, where $D_{i},S_{i}$ are
doublet (singlet) superfields, $H$ represents Higgs superfields,
and $y_{i}$ are Yukawa coupling constants.  Then, the relevant
interaction Lagrangian reads
\beq
  \CL_{\rm int} \sim 
     y_{i}^{2} \eta \sigma D_{i}^{2} S_{i}^{2} + \dots,
\eeq
which leads to the decay width $\Gamma$ given by 
\beq
  \Gamma \sim \Sigma_{i} y_{i}^4 \eta^2 m_{\sigma}^3.
  \label{eq:gamma}
\eeq
Here $m_{\sigma} = 2\sqrt{g}e^{1/g}v^2$ is the mass of $\varphi$ at
the vacuum expectation value $\eta = \sqrt{2/g}$. 
Thus the universe is reheated even if the system of $\Phi$ and $X$
has no direct coupling with the standard fields, and
the reheat
temperature $T_{R}$ is given by
\beq
  T_{R} \sim 0.1 \bar{y} \eta m_{\varphi}^{3/2},
\eeq
where $\bar{y}=\sqrt{\Sigma_{i} y_{i}^4}$. Taking
$\bar{y} \sim 1$, the reheating temperature $T_{R}$ is given by
\beq
  T_{R} \sim 10^{-10} - 10^{-8},
\eeq
for $0.02 \le c \le 0.1$, which is low enough to avoid the
overproduction of gravitinos in a wide range of the gravitino mass
\cite{Ellis}.
  
Finally we comment on higher-order terms in the K\"ahler potential.
Fourth order terms in the K\"ahler potential such as
\beq
  \Delta K = \frac{k_{1}}{2}|X|^2 (\Phi+\Phi^{\ast})^2 
        +\frac{k_{2}}{4}|X|^4 +\frac{k_{3}}{12}(\Phi+\Phi^{\ast})^4,
\eeq
roughly generate three minor changes associated with $k_{i}$.  The
first change is as follows: during chaotic inflation, the mass squared
$m_{X}^{2}$ of $X$ acquired the additional terms
\bea
  m_{X}^{2} &\simeq& 2 g^{2} v^{4} \chi^{2} 
                   - \frac{k_{2}}{4} g^{2} v^{4} \chi^{4} \non \\
            &\simeq& -3k_{2} H^{2}.
\eea
Thus if $k_{2} < -\frac34$, $X$ oscillates rapidly so that the
amplitude of the oscillation damped into zero. Thus we can safely set
$|X|$ to be zero in the whole analysis of the present model. The
second one is the change of $ c = 2(g-1)$ into $ c = 2(g+k_{1}-1)$.
$k_3$ is almost irrelevant for the dynamics of $\varphi$ field and
only changes its vacuum expectation value $\eta$ due to the
redefinition of $\varphi$ with a canonical kinetic term.

Furthermore, the inflaton may have interactions with standard light
particles $\psi_{i}$ in the K\"ahler potential which is invariant
under the Nambu-Goldstone-like shift symmetry and $U(1)_{R} \times
Z_{2}$ symmetry, $K(\Phi,\psi_{i}) = \Sigma_{i} (\lambda_{i}/2)
(\Phi+\Phi^{\ast})^2 |\psi_{i}|^2$. Then, the interaction Lagrangian
density is given by $\CL_{\rm int} = \Sigma_{i} \lambda_{i} \varphi^2
\del^{\mu}\psi_{i}\del_{\mu}\psi_{i}^{\ast}$, which yields the decay
width $\Gamma$ of the same order as that in Eq. (\ref{eq:gamma}) and
gives the similar reheating temperature.

\section{Conclusions and Discussions}

\label{sec:conclusions}

In the present paper, we have proposed a consistent scenario of
successive inflation in supergravity.  Chaotic inflation takes place
around the Planck scale so that the universe can live long enough. In
this regime, the inflaton responsible for new inflation dynamically
relaxes toward zero so that new inflation sets in. The reheating
temperature is low enough to avoid overproduction of gravitinos in a
wide range of the gravitino mass. Furthermore, our model generally
predicts a tilted spectrum with the spectral index $n_{s} < 1$, which
may naturally explain the recent observation of anisotropies of CMB by
the BOOMERANG experiment \cite{BOOMERANG} and the MAXIMA experiment
\cite{MAXIMA}.  Our model can also accommodate a leptogenesis scenario
where the inflaton decay produces heavy Majorana neutrinos and in
succession these neutrinos decay to generate the lepton asymmetry to
explain the baryon asymmetry observed in our universe \cite{Asaka}.

In the present model, the initial value of the inflaton for new
inflation may be so close to the local maximum of the potential that
the universe enters a self-regenerating stage.  Therefore all the
scales observable today left the Hubble radius during the last
inflation and we cannot verify the chaotic inflation stage directly
because the minimum of K\"ahler potential during chaotic inflation
coincides with the local maximum of the potential for new inflation.
By appropriately shifting the local minimum of the new inflaton's
potential during chaotic inflation one can construct a model in which
duration of new inflation is short enough that the trace of chaotic
inflation is observable on the large-scale structure.  This issue will
be discussed in a forthcoming publication.

\subsection*{ACKNOWLEDGMENTS}

M.Y. is grateful to M. Kawasaki and T. Yanagida for useful
discussions. M.Y. is partially supported by the Japanese Society for
the Promotion of Science. J.Y. is supported in part by the Monbusho
Grant-in-Aid, Priority Area ``Supersymmetry and Unified Theory of
Elementary Particles''(\#707) and the Monbusho Grant-in-Aid for
Scientific Research No.\ 11740146.

\end{document}